\documentclass[11pt]{book}
\usepackage{Wiley-AuthoringTemplate}
\usepackage{chapterbib}
\usepackage{setspace}
\usepackage[sectionbib,numbers]{natbib}

\usepackage[version=4]{mhchem}

\definecolor{darkorange}{RGB}{179,98,0}

\makeindex

\begin{document}

\frontmatter

\mainmatter

\chapter[Ch1 Origins of Life on Exoplanets]{Origins of Life on Exoplanets}

\author*[1,2,3]{Paul B. Rimmer}

\address[1]{\orgdiv{Department of Earth Sciences}, 
\orgname{University of Cambridge}, 
\street{Downing St}, \city{Cambridge}, \postcode{CB2 3EQ}, \country{United Kingdom}}%

\address[2]{\orgdiv{Astrophysics Group Cavendish Laboratory}, 
\orgname{University of Cambridge}, 
\street{JJ Thomson Ave}, \city{Cambridge}, \postcode{CB3 0HE}, \country{United Kingdom}}%

\address[3]{\orgdiv{MRC Laboratory of Molecular Biology}, 
\orgname{Cambridge Biomedical Campus}, 
\street{Francis Crick Ave}, \city{Cambridge}, \postcode{CB2 0QH}, \country{United Kingdom}}%

\newcommand{\gunter}{W\"{a}chtersh\"{a}user}

\address*{\email{pbr27@cam.ac.uk}}

\maketitle

\begin{abstract}{Abstract}
\singlespacing
I show that exoplanets can be used to test origins scenarios. Origins scenarios start with certain initial conditions, proceed via a network of chemical reactions and, if successful, result in a chemistry that is closer to a living system than the initial conditions. Exoplanet environments can be applied to test each of these three aspects of origins scenarios. I show what tests can be applied to the UV-driven cyanosulfidic scenario and how the application of some of these tests has already falsified certain versions of this scenario. Testing initial conditions has replaced certain reactants with others and has affected the overall chemical network underlying the cyanosulfidic scenario. The sequence of reactions the scenario invokes provide a predicted upper limit on the ubiquity of life in the universe that has ample room for improvement. The outcome of the experiments in different environments is part of a predicted distribution of biosignature detections that can be compared to future observed distributions.
\end{abstract}


\begin{abstract}{Author Links}\\
\url{http://www.mrao.cam.ac.uk/~pbr27}\\
\url{https://scholar.google.co.uk/citations?user=BBcDP-0AAAAJ}
\end{abstract}

\singlespacing

\section{Introduction}
\label{sec:intro}

\begin{quotation}
``Anybody who thinks they know the solution to this problem (of the origins of life) is deluded; but anybody who thinks this is an insoluble problem is also deluded.'' -- Leslie Orgel\footnote{Related by Bardi, J. S. in {\it News and Views} Vol. 3 (The Scripps Research Institute, 2004).}
\end{quotation}

Since the classic experiment of Miller \cite{Miller1953}, thousands of experiments have been performed and computational models run, and dozens of scenarios have been proposed to address the question of how life originated. Some of these scenarios concentrate on a handful of promising reactions, starting from pure mixtures with precisely controlled conditions \citep{Sanchez1970}. Others start from interesting reactions embedded in biological or geochemical systems or both \citep{Cody2001}, trying to work backwards to discover the bridge between the two, with some promising results \citep{Lane2012}. Still others are inspired by the myriad compounds discovered in interstellar space near our galactic center \citep{McGuire2016}, or in the ices of comets \citep{Altwegg2016}. Some approaches seek to construct a plausible story of how life arose on Earth \citep{Damer2015,Patel2015}, while others instead search for the fundamental principles that govern life \citep{Cronin2016}, if any are to be found. 

All scenarios and experiments are joined to early Earth by analogy, though some analogies are much closer to the real thing (e.g., Milshteyn et al. \citep{Milshteyn2018}). The use of analogy raises a serious problem for origins research and opens the entire field up to a flood of criticisms, well-summarized in a book review by William Bains \citep{Bains2020}. Underlying these criticisms is a methodological question: how will we ever know if we found the answer? What hints are there that we are even on the right track? The central methodological issue in origins research is testability.

This issue was anticipated early on by G\"{u}nter \gunter, who applied the principle of falsifiability and the link between testability and informativeness of a theory, following Karl Popper \citep{Popper1959}, to construct his scenario and to provide a clear and useful template for the development of alternative scenarios \citep{Wachtershauser1988}. Testability is difficult to achieve, however, since the lab environment is artificial and all geochemical environments yet discovered on Earth where life can survive have been transformed by the presence of life.

In this chapter, I propose that exoplanets can act as new laboratories on which to test origins scenarios. These tests will require an older style of experiment. Though Miller's experiment showed up so early on the scene that we could not understand the results or whether the experimental conditions themselves were genuinely plausible, Miller's experimental approach is exactly what the chemist should return to in the next few years, not blindly, but with decades of chemical knowledge at our disposal. We can construct more realistic, more messy, environments within which to perform experiments (see for example \citep{Moller2017}). These experiments will doubtless lead to the discovery of novel scenarios. Hopefully they will also rule out others.

Applying exoplanet science to test origins hypotheses will require a different way of thinking about specific scenarios, not just in one environment on Earth, but displaced to a wide range of diverse extraterrestrial environments, where the chemistry will be found to succeed or fail, or to reveal an unexpected new path to explore. This approach can be helped, I find, by a shift in language.

\section{How to Test Origins Hypotheses}
\label{sec:test}

An intermediate goal for origins of life research is the discovery of a {\it scenario} where some local and available chemical and physical conditions lead to a series of chemical reactions that proceed without the assistance of a chemist to a protocell: a self-contained self-replicating metabolizing macromolecular machine with the potential to evolve into something like the last universal common ancestor (LUCA) via random mutations and natural selection.

All proposed scenarios are far from this intermediate goal. Nevertheless, any given scenario can be criticized on the basis of this desired outcome. Maybe a scenario has implausible initial conditions. Or maybe the proposed sequence of reactions are unlikely to proceed without help. Finally, since the outcome is a long way from a protocell, and may look quite different from a functional protocell, let alone LUCA, it could be argued that the outcome will turn out to be irrelevant. These three criticisms unique to origins of life are on the basis of:
\begin{enumerate}
\item Initial conditions.
\item Sequence of reactions
\item Outcome
\end{enumerate}
The underlying problem is that when someone tries an experiment in the lab and it works, this only shows that the experiment will work in the lab. Granted, if outside the lab the same conditions and sequence of events just happen to occur, the outcome will be the same. But the chances of all these events happening in a way identical to a lab and without the chemist will take on the character of a Rube Goldberg machine. But this requires too much specificity. Some decisions are made for experimental convenience, some for the sake of characterizing the product, and some decisions are arbitrary. Many changes will do nothing to the chemistry. Some changes will result in a different outcome, but a different outcome is not the same as failure. Some changes will likely improve the chemistry, and many other changes will likely disrupt the chemistry \citep{Walton2020}. 

I find it useful to think about these components of a scenario using the language of ‘natures’ and ‘powers’ (following Cartwright \citep{Cartwright1999}\footnote{This language has a long history in origins science, since natures and powers are closely related to propensities \citep{Cartwright2008}, and propensities were central to the Popperian cosmology \citep{Popper1959b}, used to great effect by \gunter \citep{Wachtershauser1988}.}). By the {\it nature} of a thing, I mean what a thing tends to do in its immediate environment. This is a result of the interactions between the powers of the thing itself and its environment. The nature of a molecule can be ascertained within a lab by series of experiments across a wide range of physical and chemical conditions, and the context of that understanding can be broadened by setting up lab conditions to more closely approximate plausible natural environments, and by exploring the environmental parameter space to find out what environmental factors matter for the chemistry. 

We can then apply our knowledge of the nature of these molecular species to model their behavior within a wider chemical context, and can predict both the intermediate and immediate consequences of that behavior and the interaction of these molecular species and their environment. Some of these immediate consequences may be preserved in the environment, and may be observable. For instance, there has been significant progress toward predicting outcomes of prebiotic chemical scenarios that can then be tested against observations by Mars 2020 in Jezero Crater \citep{Sasselov2020}.

The criticisms and problems now become tests:
\begin{enumerate}
\item Does the environment invoked by the scenario provide the necessary initial conditions for the reactions to take place?
\item What is the probability that these reactions will take place in sequence and unaided in this environment?
\item How likely is it that the outcome of these reactions will result in a protocell? What paths are opened up for chemical progress and what paths are closed off, either by the environment or by the direct outcome of the reaction sequence?
\end{enumerate}
Here I will focus on rocky planetary\footnote{{\it Planetary} here includes some satellites of other planets, like Enceladus, Europa and Titan.} environments. These environments can be split up into Earth environments and other planetary environments within our solar system (Mars, Venus, certain moons) and outside our solar system (exoplanets). The two relevant factors that distinguish between these groups is {\bf resolution} and {\bf population size}.

{\bf Earth} is the planet where we have by far the most information and best resolution. This resolution makes clear that Earth is not a single environment, but many, spanning a variety of temperatures from below zero to thousands of degrees Celsius, pressures much less than a bar to many GPa, and a variety of redox conditions from highly oxidizing to highly reducing \citep{Etiope2004}. Nonetheless, Earth is a sample size of one, and all environments relevant for prebiotic chemistry have been transformed by life.

{\bf Rocky Planetary Bodies in the Solar System.} The resolution we have for planetary bodies in the solar system varies with distance and ease of investigation \citep{Kane2019}. The sample size where we have decent resolution includes a dozen or so objects. Some of these objects, perhaps all, are and have always been lifeless (but see \citep{Chyba2001,Greaves2020}). Still, it is challenging to apply statistics usefully to such a small sample size, and the sample size is biased since all these rocky bodies are part of the same single solar system.

{\bf Exoplanets} are the planets we know least about. Most exoplanets we've discovered are inferred from the changing light of their host star, itself a small fraction of a single pixel on a detector. However, the number of exoplanets is large and rapidly increasing, spanning many stellar systems around many different types of stars. The sample is shaped by observational biases, but new detection techniques will slowly mitigate these.

This chapter will focus on how exoplanets can inform tests of origins scenarios. In the next section, I will briefly discuss exoplanet detection and characterization.

\section{Exoplanets as Laboratories}
\label{sec:exo}

\begin{figure}
\centering
\includegraphics[width=1.1\linewidth]{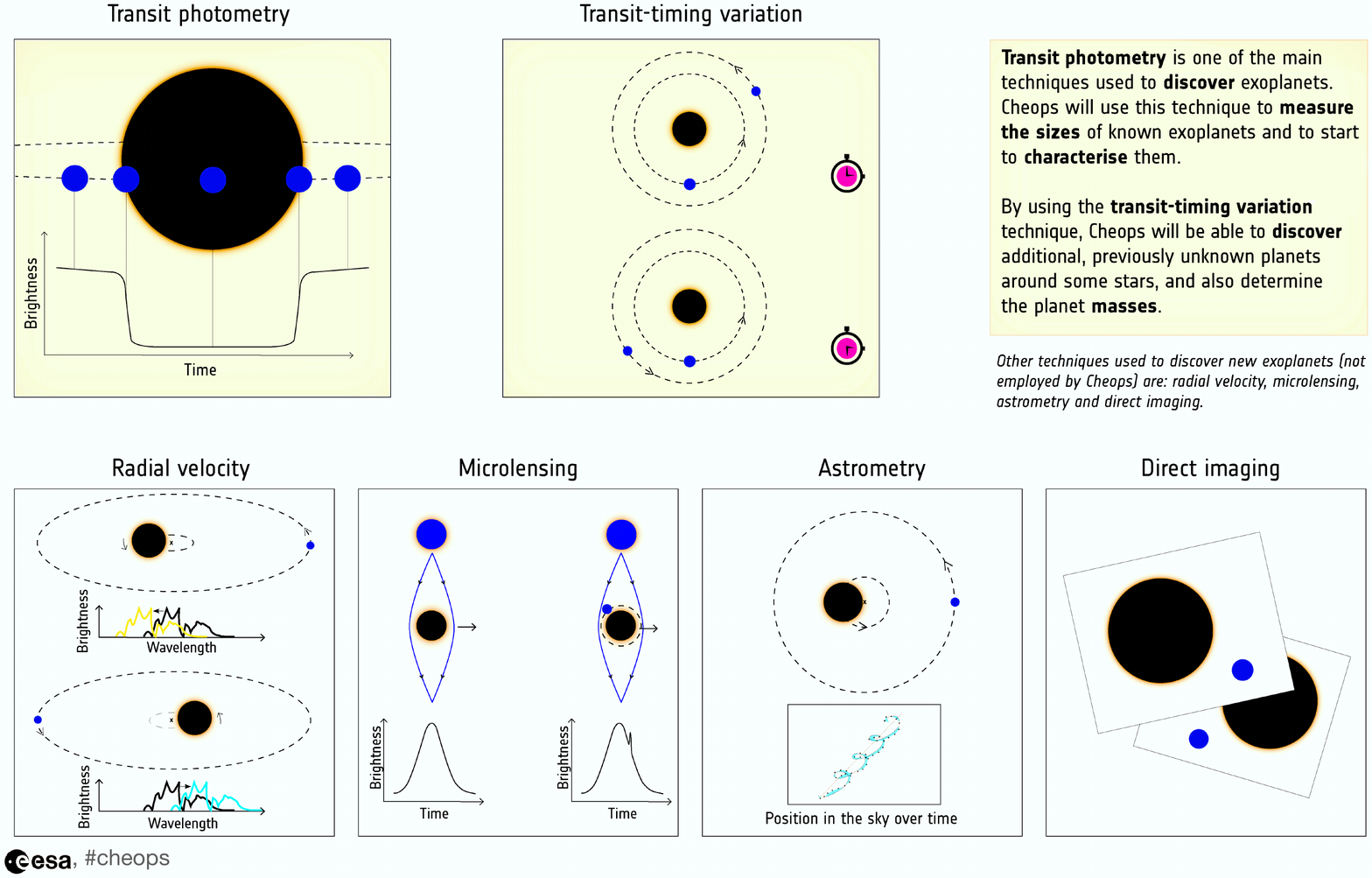}
\caption{Schematic of a variety of detection methods, modified from an infographic from the European Space Agency for the Cheops mission (\url{https://www.esa.int/ESA_Multimedia/Sets/Exoplanets_infographics/(result_type)/images}).
\label{fig:exoplanets}}
\end{figure}

Exoplanets are planets that orbit stars other than the Sun. Thousands of exoplanets have been discovered\footnote{See \url{http://exoplanet.eu/}}. Exoplanets orbit a wide variety of stars and their properties span a wide range, e.g. distance from their host star, their sizes and masses. Exoplanets are primarily discovered by three different techniques (see Figure \ref{fig:exoplanets}, and see \citep{Wright2013} for a detailed review):
\begin{enumerate}
\item {\bf Transit:} Where the star dims as the planet passes in between the star and the observer.
\item {\bf Radial Velocity:} Where the planet's gravitational pull on the star causes the light of the star to shift in frequency
\item {\bf Direct Imaging:} Where the observer blocks the light from the star in order to observe the light of the planet directly.
\end{enumerate}
Transits determine the ratio of the planet's radius to the star's radius. Where the star's radius is known, the size of the planet can be determined. Radial velocity and knowledge of the star's mass can similarly be used to obtain the mass of the planet. The combination of these techniques can constrain the planet's density, giving us an idea about whether a planet is a rocky planet or a gas planet \citep{Seager2007,Zeng2016}. The periodic signal of the transit or radial velocity, or the measured distance between a star and an orbiting directly imaged planet gives the planet's distance from its host star. The temperature and brightness of the star, often expressed in terms of its stellar type, can be used to delineate a range of distances from the star within which liquid water can exist stably on a planet's surface \citep{Kasting1993}. We can call this the {\it habitable zone}.

One way the atmosphere of a planet can be characterized is when the planet's transit is observed at different wavelengths. The apparent radius of the planet will change with wavelength depending on the absorption properties of its atmosphere. This task is much easier to accomplish around small stars, where the size of the atmosphere divide by the size of the star is larger \citep{Segura2005}, and where the habitable zone is nearer to the star, so transits of planets within the habitable zone are more frequent. We do not yet have the observational capabilities to observe atmospheres of rocky planets within the habitable zones even of ultra-cool stars.

In this chapter, I will focus on rocky planets within the habitable zones of their host stars (see Kane et al. \citep{Kane2016} for a catalogue of these planets). But first, I will briefly say something about the planets whose atmospheres we can characterize today. 

\paragraph{Gas Planets} Gas giant planets ($R_p \gtrsim 0.5 R_J$) are by far the easiest planets to characterize. Intrinsically interesting, gas giants are also relevant for origins questions in at least two ways. {\bf (a)} Their composition gives insight into planet formation history and disk chemistry \citep{Qi2013,Madhu2017}, and {\bf (b)} physical processes, such as atmospheric photochemistry, stellar energetic particle chemistry, and impact-induced chemistry, can be explored on these exoplanets.

\section{The Scenario}
\label{sec:scenario}

I will discuss the tests listed in Section \ref{sec:test} for only one scenario. These tests can in principle be applied to any other sufficiently specific scenario.

The scenario I discuss is the cyanosulfidic scenario. It began when a 1 km sized impactor struck the surface of a volcanic island on early Earth, during the time it had a transient reducing atmosphere \citep{Genda2017,Ritson2020}. That impact delivered phosphorus in the form of phosphate (\ce{PO_4^{2-}}) \citep{Patel2015}, and generated a crater with large amounts of hydrogen cyanide (\ce{HCN}, \citep{Ferus2017}). This \ce{HCN} rained down and reacted with available Fe(II) to form ferrocyanide (\ce{[Fe(CN)_6]^{4-}}). The volcanism heated the ferrocyanide in the crater causing it to form various organometallic and other complexes, depending on the local mineralogy and temperature (see Table \ref{tab:organometal}).

\begin{figure}
\centering
\includegraphics[trim={0cm 0.5cm 0cm 0.5cm},clip,width=1.1\linewidth]{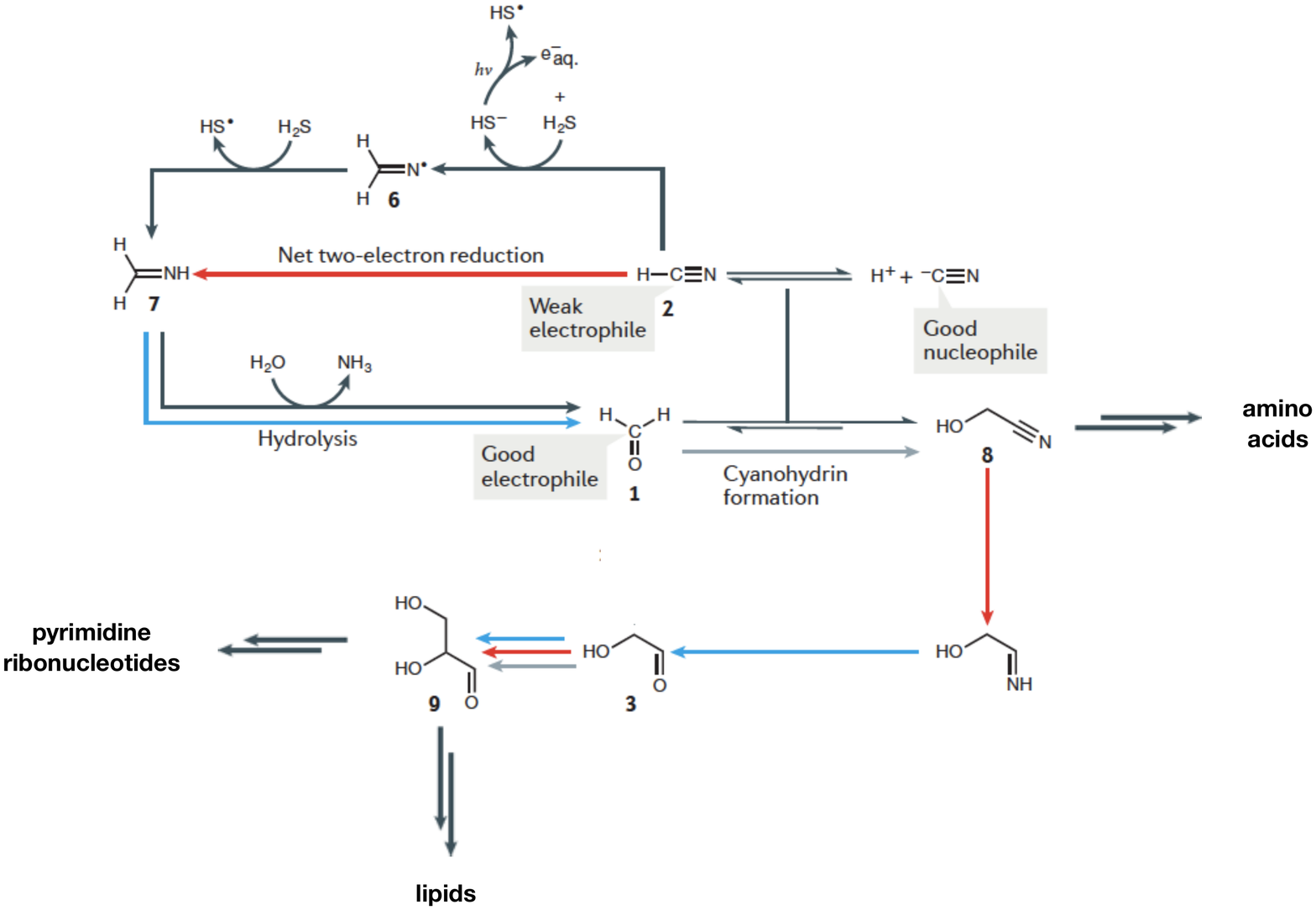}\\
\includegraphics[trim={5cm 5cm 3cm 5cm},clip,width=\linewidth]{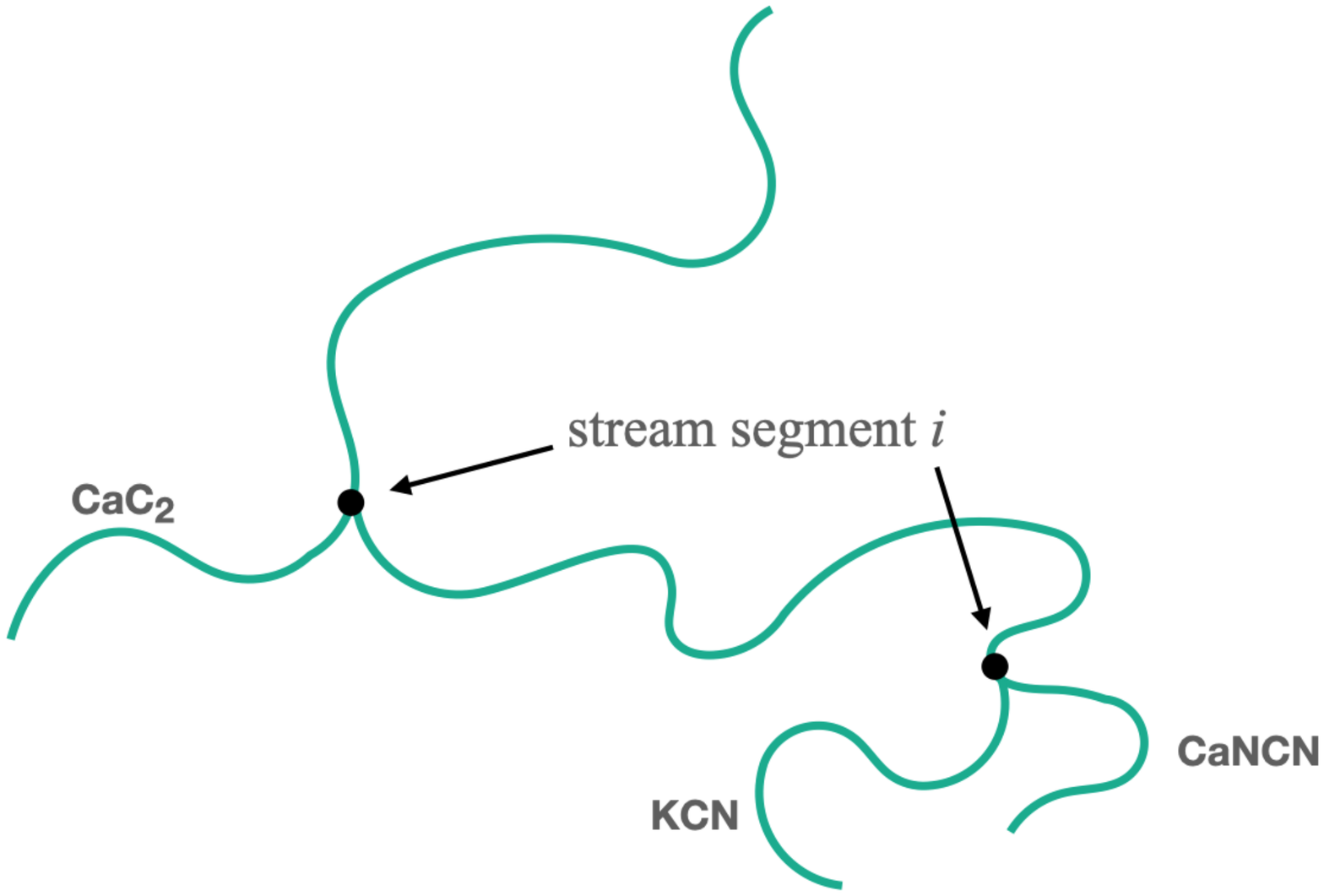}
\caption{{\bf (Top)}. First steps of the cyanosulfidic photochemical scenario, adapted from Sutherland \citep{Sutherland2017} with permission. {\bf (Bottom)}. A series of streams intersecting each other, running over different organometallic feedstock compounds. A stream segment $i$ is identified, relevant for Section \ref{sec:prob}.
\label{fig:scenario}}
\end{figure}

Small streams of water flowed over this material, releasing volatile compounds (see again Table \ref{tab:organometal}) and initiated a series of prebiotic chemical reactions as discussed in detail by Patel \citep{Patel2015}, generating small sugars that proceeded to form amino acids, phospholipid precursors and nucleotides, including pyrimidine ribonucleotides \citep{Patel2015}, and purine deoxyribonucleotides \citep{Xu2020}, as shown in Figure \ref{fig:scenario}. Subsequent nitrate chemistry generated non-enzymatic activating agents \citep{Mariani2018}, that facilitated the oligomerization of these compounds \citep{Liu2019}.

The different sequences of reactions are not all compatible with each other, so different streams must have carried different reactants and intermediate products. These streams must have intersected at the right time and in the right sequence for the scenario to succeed. See again Figure \ref{fig:scenario}.

I choose this scenario for several reasons. First, it seems to be the most plausible scenario yet discovered for producing precursors to RNA, DNA, proteins and membranes in the same local environment. On this point, see Chapter {\bf X} by Sutherland. Second, the scenario has been given in sufficient detail \citep{Patel2015,Xu2018,Mariani2018,Ranjan2016,Sasselov2020,Rimmer2020} to apply the above tests and to potentially falsify aspects of the scenario. There may be other geochemical environments where the same sequence of chemical reactions can occur, e.g. hydrothermal vents \citep{Rimmer2019}, but these alternatives have not yet been worked out in sufficient detail to apply the above tests.

Finally, I have a pragmatic reason for choosing this scenario. The scenario depends on starlight, and starlight is the thing about exoplanet systems we know the most about. \textbf{For the next few decades, any scenario that uses starlight will be the most straightforward to connect to exoplanet observations.}

\begin{table}
\caption{Organometallic Feedstock Compounds and their Generation \label{tab:organometal}}{%
\begin{tabular}{@{}cccc@{}}
\toprule
Starting Compound            & Transformed Compound                       & Temperature    & Other Species/Conditions\\
\midrule
\ce{(Fe,Ni)_3P}              & \ce{(Fe,Ni)_3(PO_4)_{2}}$\cdot$\ce{8 H_2O} & RT\footnotemark[1]             & Liquid Water            \\
HCN                          & \ce{\{(Na,K)_4,Mg_2,Ca_2\}[Fe(CN)_6]}          & RT             & \ce{Fe(II)}\footnotemark[2]  \\
\ce{(Na,K)_4[Fe(CN)_6]}      & \ce{(Na,K)CN}                              & 700$^{\circ}$C &   \\
\ce{Mg_2[Fe(CN)_6]}          & \ce{Mg_3N_2}                               & 420$^{\circ}$C &   \\
\ce{Ca_2[Fe(CN)_6]}          & \ce{CaNCN}                                 & 660$^{\circ}$C &   \\
\ce{CaNCN}                   & \ce{CaC_2}                                 &1000$^{\circ}$C &   \\
\ce{(Na,K)CN}                & \ce{HCN}                                   & RT             & Liquid Water  \\
\ce{Mg_3N_2}                 & \ce{NH_3}                                  & RT             & Liquid Water  \\
\ce{CaNCN}                   & \ce{CN_2H_2}                               & RT             & Liquid Water  \\
\ce{CaC_2}                   & \ce{HC_3N}                                 & RT             & Liquid Water, \ce{Cu(II)}, \ce{HCN}  \\
\botrule
\end{tabular}}{\footnotetext[]{Based on Figure 2 from Patel et al. \citep{Patel2015}}
\footnotetext[1]{Room Temperature.}
\footnotetext[2]{Best if \ce{Fe(II)} is aqueous. The outcome depends on the content of Mg vs Na and K available within the basalt and the presence of surface metal oxides, hydroxides, chlorides, carbonates, etc.}}
\end{table}

\section{Initial Conditions}
\label{sec:initial}

The first test for which exoplanets can act as laboratories is as a test for initial conditions. What initial conditions are required for the success of the cyanosulfidic scenario, and what fraction and distribution of exoplanets meets these conditions? This question can be addressed both in the lab and by future observations of exoplanet atmospheres. Initial conditions can be split into two groups: chemical initial conditions and physical initial conditions.

\subsection{Chemical Initial Conditions}
\label{sec:initial-chemical}

The chemical initial conditions for the cyanosulfidic scenario are various, including cyanoacetylene, cyanamide, and phosphate. I will concentrate on the two initial chemical conditions that give the cyanosulfidic scenario its name. The cyanosulfidic scenario depends on both cyanide and sulfide, and I will discuss the availability of each of these species in turn.

\subsubsection{Hydrogen Cyanide}
\label{sec:initial-chemical-hcn}

Hydrogen cyanide (\ce{HCN}) can be formed in large quantities when there is available free nitrogen in a local environment where the C/O ratio is $\geq 1$. \ce{HCN} can be formed thermochemically, from lightning \citep{Chameides1981,Ardaseva2017}, or in magmas \citep{Rimmer2019}, and can be formed photochemically when thermospheric \ce{N_2} is photodissociated and the resulting atomic nitrogen diffuses downward, where it interacts with the dissociation products of \ce{CH_4} or of \ce{CO} and \ce{H_2} \citep{Rimmer2019b}. The \ce{N_2} can be dissociated deeper in the atmosphere by stellar energetic particles \citep{Airapetian2016}.

\ce{HCN} can also be delivered by smaller impactors, either primordial \citep{Oberg2015}, or from the destruction of delivered reduced nitrogenous compounds during the impact \citep{Todd2020}. Large impactors carry \ce{Fe^{0+}}, which can be oxidized on impact by \ce{H_2O} and \ce{CO_2}, resulting in \ce{FeO}, \ce{CO} and \ce{H_2}. The heat from the impact can allow these species to further reduce other atmospheric constituents, resulting in large amounts of \ce{CH_4} and \ce{NH_3}. This transient ``warm Titan'' atmosphere can persist for millions of years \citep{Zahnle2020}. Subsequent impacts within this transient atmosphere produce large amounts of \ce{HCN} and acetylene (\ce{C_2H_2}). The \ce{C_2H_2} is in principle detectable in the atmospheres of rocky exoplanets \citep{Rimmer2020}. The search for acetylene in the atmospheres of young rocky exoplanets can help us constrain the fraction of planets that have had these atmospheres which are very useful for prebiotic chemistry \citep{Benner2020}.

Even without impacts, \ce{HCN} can be produced in large quantities, but only given sufficiently reducing local or global environments. The relative ubiquity of these environments can be constrained by searches for atmospheric ``prebiosignatures' such as \ce{CO}, \ce{CH_4}, \ce{HCN}, and \ce{C_2H_6} on exoplanets \citep{Rugheimer2020}, species that are either themselves feedstock molecules for prebiotic chemistry or are tracers of reduced atmospheres where production of these feedstock molecules is most conducive.  

\begin{figure}
\centering
\includegraphics[width=\linewidth]{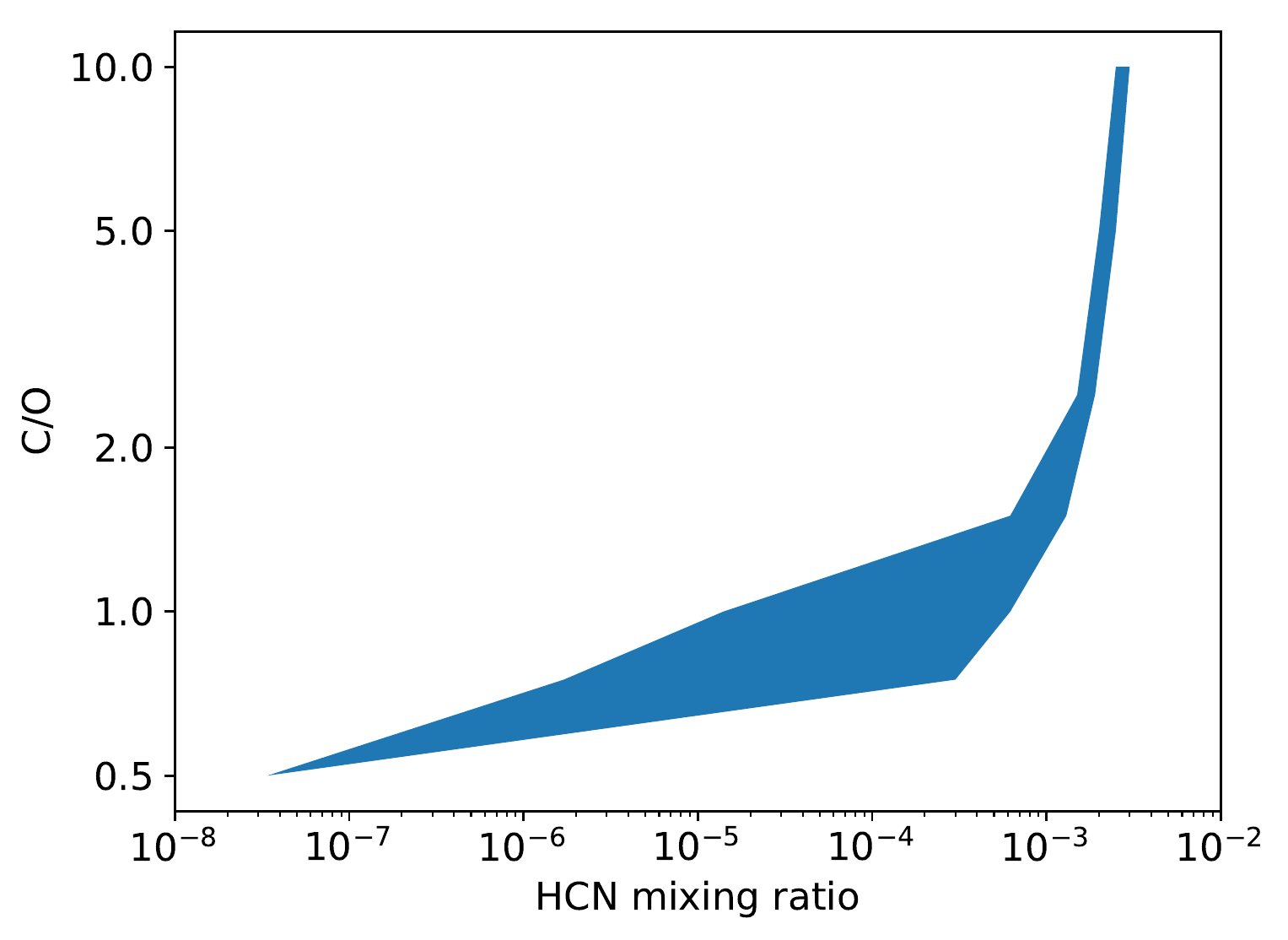}
\caption{The photochemical upper atmospheric mixing ratio of \ce{HCN} based on the local atmospheric C/O ratio. Taken from Rimmer \& Rugheimer \citep{Rimmer2019b}.
\label{fig:hcn}}
\end{figure}

\subsubsection{Sulfite and Sulfide}
\label{sec:initial-chemical-sulfur}

The cyanosulfidic scenario as presented by Patel et al. begins with \ce{H_2S} and \ce{HCN} in liquid water with phosphate and UV light. At all degassing pressures with plausible early Earth magma temperature and mantle redox, \ce{SO_2} degassing exceeds \ce{H_2S} degassing by at least a factor of 2 \citep{Gaillard2014}. In addition, the pKa of \ce{H_2S} is 7 and the pKa of \ce{SO_2} is 1.81, so for acidic to neutral waters, \ce{SO_2} is much more soluble than \ce{H_2S}. Ranjan et al. \citep{Ranjan2018}, first noticed this issue and further determined that aqueous \ce{SO_2} will be converted primarily into sulfites: \ce{HSO_3^-} and \ce{SO_3^{2-}}. Like sulfides, sulfites can undergo electron photodetachment under UV light, and the resulting chemistry is similar \citep{Xu2018}. Both \ce{H_2S} and \ce{SO_2} are plausible for any volcanic environment on early Earth, but \ce{H_2S} will not plausibly be present without \ce{SO_2}, and not in excess of \ce{SO_2}, except in alkaline environments.

\subsection{Physical Initial Conditions}
\label{sec:initial-physical}

This scenario also requires a range of physical conditions that can be tested by future experiments and exoplanet observations. Requirements include:
\begin{enumerate}
\item Surface liquid water
\item Dry land
\item Sufficient ultraviolet(UV) flux
\end{enumerate}
Other physical conditions such as volcanism and a history of impacts are also required in the case that they may be required to provide sufficient feedstock molecules (see the above Section \ref{sec:initial-chemical}), but are not listed here because there may be alternative ways to obtain the required chemical initial conditions.

The first requirement is accounted for by the Liquid Water Habitable Zone (see Section \ref{sec:exo}). The second requirement can be tested by looking at water worlds, planets that have a sufficiently large mass fraction of water that they have no dry land. Alternative scenarios may or may not work on water world planets \citep{Kite2018}, but the scenario I consider very likely cannot. Water worlds can be identified by constraining planetary density and by atmospheric observation (Section \ref{sec:exo}).

\begin{figure}
\centering
\includegraphics[width=\linewidth]{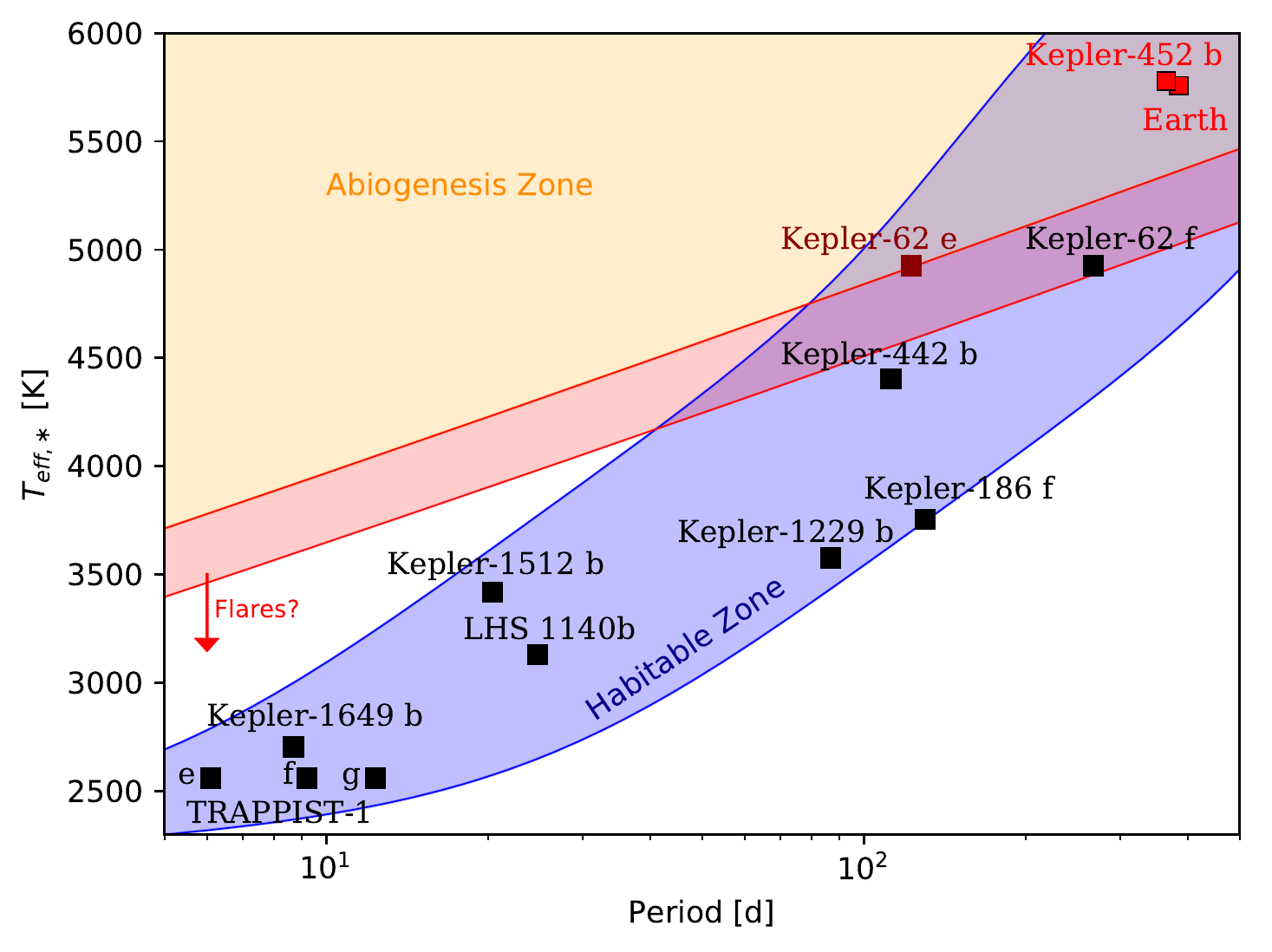}
\caption{The abiogenesis zone (orange, with error bars in red) and liquid water habitable zone (blue) as a function of planetary period and stellar effective temperature. Outside the liquid water habitable zone, planets with surface temperatures regulated by \ce{CO_2} and \ce{H_2O} greenhouse gases cannot sustain liquid water on their surfaces. Outside the abiogenesis zone, the quiescent UV emission of the star is not intense enough to allow for the prebiotic photochemical scenario presented in this chapter. Figure taken from Rimmer et al. \citep{Rimmer2018}.
\label{fig:abio-zone}}
\end{figure}

\paragraph{The Abiogenesis Zone} The involvement of UV light makes possible the delineation of a zone around a star outside of which this chemistry cannot take place. I call this the ``Abiogenesis Zone'' \citep{Rimmer2018}. The Abiogenesis Zone is based on a competition between the UV photodetachment of electrons from either \ce{SO_3^{2-}} or \ce{HS^-}, and subsequent reduction of either \ce{HCN} or glycolonitrile, versus reactions in the dark that result in inert adducts. By comparing the rates of these reactions, Rimmer et al. \citep{Rimmer2018} were able to delineate the Abiogenesis Zone for the quiescent emission of main-sequence stars. I show the Abiogenesis Zone based on \ce{SO_3^{2-}} photodetachment in Figure \ref{fig:abio-zone}.

One implication of Rimmer et al. \citep{Rimmer2018} is that early Earth lies outside the Abiogenesis Zone with regard to \ce{HS^-} photodetachment. This result falsifies the version of the cyanosulfidic scenario from Patel et al. \citep{Patel2015}, but not the version of the scenario from Xu et al. \citep{Xu2018}.

Another implication of Rimmer et al. \citep{Rimmer2018} is that ultracool stars cannot host planets on which this chemistry will work, based on the quiescent emission of those stars. This includes the vast majority of temperate rocky planets so far discovered \citep{Kane2016,Gillon2017}. Follow-up studies test whether flares may provide sufficient UV light, testing the hypothesis of Ranjan et al. \citep{Ranjan2017} and the experimental implications of Rimmer et al. \citep{Rimmer2018} against observations. These studies indicate that it is likely that the vast majority of these stars' present activity levels are insufficient for this scenario \citep{Gunther2020,Ducrot2020,Glazier2020}. The search for biosignatures on these planets will test this scenario.

\section{Chances of Success}
\label{sec:prob}

I now discuss the chances that the sequence of reactions invoked by this scenario will succeed. To this end, I return to the picture of the intersecting streams (Figure \ref{fig:scenario}). In this figure, there is a stream segment labeled $i$. The chemistry has a chance of succeeding for each segment, $c_i$. Since, for this scenario, the chemistry must be segregated into different streams, and the intersection of the streams out of sequence leads to unwanted chemistry and to a failure for the scenario, we can estimate the probability that the scenario succeeds in bringing about the origins of life, where $N$ separate stream crossings are required in the correct sequence to bring life's origin about. We don't yet know whether the entire scenario will require more and more streams, but we know this scenario at this early stage requires at least three, probably four. How many more are required will say a lot about how common or rare life is predicted to be elsewhere in the universe, given that this scenario is the most plausible scenario for life's origins.

We start with a crater full of streams, and the surface of the crater has organometallic compounds that will react in the water to release cyanide, cyanamide, acetylene, the prebiotically relevant compounds. First, the stream we follow must originate in the part of the crater with the right starting organometallic feedstock, and the reactions to take place in the first segment must succeed before the intersection with the next stream. This gives us a leading term of $c_1/N$. If a stream intersects another stream, there is a chance the stream will disrupt the chemistry or will not affect the chemistry, and the fraction of streams that do not affect the chemistry for segment $i$ is $M_i/N$.

The probability for a single stream to succeed is then the product of the probabilities for each segment, or:
\begin{equation}
P_s = \dfrac{c_1}{N}\prod_{i=2}^N\Bigg[\Big(\dfrac{1}{N-M_i}\Big)\,c_i\bigg].
\label{eqn:crossing-prob}
\end{equation}
If $M_i = N - 1$, then every stream segment is compatible with any possible stream. If $M_i = 1$, then the stream segment is compatible with no other possible stream. We assume that $M_i = 1$ for all $i$. If this were not the case and $M > 1$, then components of the chemistry could be collapsed, less streams would be needed, and the hypothesis can be effectively simplified by setting $N \rightarrow N - M$. We will also assume that all the $c_i$ are the same value for simplicity.

We can use Eq. (\ref{eqn:crossing-prob}) to estimate the probability that the chemistry will succeed, $P$, as:
\begin{equation}
P = 1 - \big(1 - P_s\big)^s,
\label{eqn:streams-prob}
\end{equation}
where $s$ is the number of streams that can flow within the crater until the reservoir organometallic feedstock is consumed. Given $\tau_{\ell}$ (y), the timescale over which a certain amount of new organometallic material is delivered or generated via impact, the timescale for life to originate will be on the order of:
\begin{align}
\tau &= \dfrac{\tau_{\ell}}{1 - \big(1 - P_s\big)^s},\label{eqn:timescale-exact}\\
P_s &= \cfrac{c^N}{N}\Bigg(\dfrac{1}{N-1}\Bigg)^{\!\!N-1}. \notag
\end{align}
When $s P_s \ll 1$ then $\big(1 - P_s\big)^s \approx 1 - s P_s$ and $N \big(N - 1\big)^{N-1} \approx N^N/e$, and Eq. (\ref{eqn:timescale-exact}) simplifies to:
\begin{equation}
\tau \approx \dfrac{1}{e} \, \dfrac{\tau_{\ell}}{s}\Bigg(\dfrac{N}{c}\Bigg)^{N}. \label{eqn:timescale-approx}
\end{equation}
We will treat $\tau$ as a function of $N$ with $s$ and $\tau_{\ell}$ as parameters that we can vary. If streams are on average 10 cm across and crater lakes are on average 1 km$^2$ in area, and each stream consumes the organometallic feedstock it immediately flows over, then 1 km$^2$ of crater lake area allows for $\sim 50000$ streams. We also assume that the total area of crater lakes is the same as the total area of lakes today, which gives us $4.2 \times 10^6$ km$^2$ of crater lake area \citep{Downing2006}, or $2 \times 10^{11}$ streams available during late accretion, over $10^7$ years. We will consider 1000 year increments, during which we estimate 420 km$^2$ of new impact-generated organometallic material is provided for the streams to flow over., and so $s = 2.1 \times 10^7$.

\begin{figure}
\centering
\includegraphics[width=\linewidth]{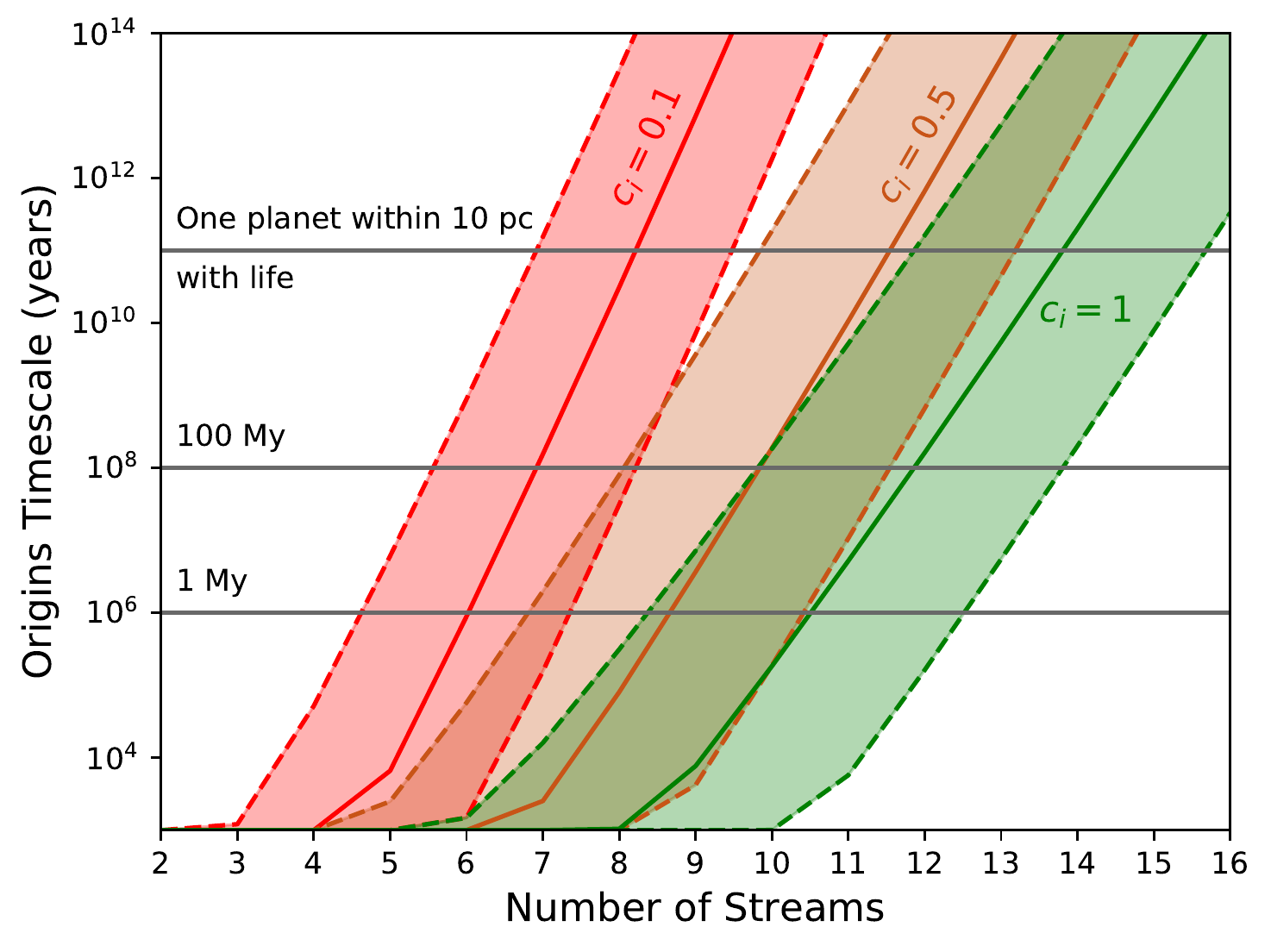}
\caption{The origins timescale (y), the timescale to start life (as defined in Section \ref{sec:prob}), vs. the number of stream crossings required, for three different probabilities of success for chemical reactions within a stream segment, $c_i = 0.1$ (red), 0.5 (orange) and 1.0 (green). See Figure \ref{fig:scenario} for a schematic of the stream crossings. 
\label{fig:streams}}
\end{figure}

Figure \ref{fig:streams} shows that the amount of time for life to originate, the ``origins timescale'' depends strongly on the chance that the chemistry in stream segment $i$ succeeds and the number of streams. If the chemistry is difficult, and unlikely to succeed through any given segment, so $c_i = 0.1$ for all $i$, then if the required number of streams is less than 6, life will arise within the first 1 My once the conditions are available. If the required number of streams is 7, it will take on average 100 My years for life to arise. If, as we assume above, the impact-generation of environments provides a $\sim 100$ My timescale for success, and if the number of streams is greater than 8, and if this scenario is the most plausible scenario for origins of life, then Earth is likely the only planet that hosts life within 10 pc.

Things become more optimistic if the probability of chemical success becomes greater. If $c_i = 0.5$ for all $i$, then the limit becomes greater than 11 streams before we predict that life is rare on this scenario. Even if the chemistry always worked out perfectly in every stream segment, such that $c_i = 1$ for all $i$, there is still a limit of 13 streams.

All of this is under the assumption of 1000 year timescales with organometallic feedstock enough for $2.1 \times 10^7$ new streams per timescale. Perhaps these estimates are off. The timescale estimate does not affect the results much at all, simply shifting the lines vertically. The number of streams matters more, but even increasing or decreasing the number of available streams, $s$, by $10^3$ shifts the line only by one or two required streams, $N$. This shift by $10^3$ is what the dotted lines around the solid lines illustrate in Figure \ref{fig:streams}.

Calculations like these could be applied to a wide range of scenarios, if the scenarios are not too vague. These calculations can be supported by real numbers by simulating sequences of reactions in prebiotic chemical scenarios and keeping track of outcomes under different environmental conditions. This would allow us to estimate the values of $c_i$, $s$ and $\tau_{\ell}$, or whatever equivalent values would apply to an alternate scenario, and to test whether the required number of streams (or other chemical transitions) is really $N$. The value of $N$ cannot be known until at least the {\it intermediate problem} (see Section \ref{sec:test}) is solved, but a lower limit can be obtained by the combination of more realistic experiments and new discoveries of subsequent chemical steps, as the scenario moves incrementally closer to solving the {\it intermediate problem}.

The resulting probabilities for one scenario can be compared to the probabilities of others given different global and local environmental conditions in order to estimate which scenario is more plausible given those conditions. These probabilities can also be applied to a Bayesian analysis of biosignatures \citep{Catling2018,Walker2018}, depending on the relevance of the outcomes of the scenarios and whether their chemical outcomes are indeed closer to life than were the initial conditions.

\section{Relevance of the Outcome}
\label{sec:outcome}

Testing the relevance of an origins scenario in terms of its chemical outcome can proceed in the lab and through future exoplanet observations.

I will say little about how this can proceed in the lab, because this direction crosses over from chemistry to biochemistry, and biochemistry is well outside my area of expertise. I can provide one case for how this test can proceed involving my work with biochemists. Methyl isocyanide is an effective non-enzymatic activating agent for amino acids, nucleotides and phospholipids \citep{Liu2019}. Methyl isocyanide can be synthesized using nitroprusside, and nitroprosside is a product of ferrocyanide and nitrites in water irradiated by a mercury arc lamp. Nitroprusside is not produced from nitrates instead of nitrites \citep{Mariani2018}. It turns out that nitrites are not nearly as plausible as nitrates in early Earth aqueous environments \citep{Ranjan2019}. However, if a more realistic broad-band light source is used, then ferrocyanide and nitrates will produce nitroprusside \citep{Rimmer2020b}, and this pathway forward is not ruled out by the environment after all.

The successful product of a future observing program to search for exoplanet biosignatures, here defined as spectroscopic signs of life, will likely be a statistical distribution of exoplanets that probably host life. The nature of that distribution as a function of parameters like stellar effective temperature and activity, the mean density of the planet, and the plausibility that it had a reducing atmosphere in its past will test and potentially falsify claims connected with certain origins scenarios.

Not much can be concluded if no life is found, apart from a vague lower limit on the difficulty for life to arise from non-life on the sample of exoplanets whose atmospheres were observed. {\it Absence of evidence for life is not evidence for the absence of life}, so applying this test will not be able to provide a specific answer that is independent of the limits of our observational capabilities. At most it would weaken the case that life is a phase transition that will inevitably be crossed within any given temperate environment. 

\section{Conclusions}
\label{sec:conclusions}

Using the cyanosulfidic scenario as an example, I show how exoplanets can be used to test origins scenarios. Initial conditions, probability of success and relevance of outcome can be tested using exoplanet environments. Present exoplanet observations, and not just of temperate exoplanets, can be applied to explore new chemical mechanisms and environments more fully in the lab, the results of which can in turn inform prior probabilities for biosignature searches. In the more distant future, exoplanet observations themselves can be used to test particular claims about given scenarios by comparing predicted to actual distributions of biosignature detections.

These tests will require a large-scale experimental exploration of the parameter space, and not just for plausible Earth environments (see Walker et al. \citep{Walker2017}). Ideally this effort will be open about what scenarios to explore. Right now there are no clear ways to falsify most scenarios presently ``on the market'', and so researchers should be free to explore their own direction as part of a friendly, supportive and tolerant research community, so long as the underlying chemistry makes sense and the scenario is being explored with an eye to future testing and potential falsification.

I suspect origins research will not require new physics. But it will require a means of testing proposed partial solutions. Different terrestrial environments have been used to perform these tests (e.g. \citep{Milshteyn2018}), and other solar system bodies have just recently come into the picture (e.g., for Mars \citep{Sasselov2020}, and Europa \citep{Vance2020}). Exoplanets are a new and largely untapped resource for this problem, if we are willing to make the effort to connect what we do in the laboratory to these alien environments in a meaningful way.


\paragraph{Acknowledgements} Some of the ideas in this chapter originated from a panel discussion at the Future of Exoplanet Research symposium sponsored by the Kavli Foundation as part of the TESS I conference. I'm grateful to Alexander Bird for helpful feedback on my Chapter's philosophical claims. I thank Ram Krishnamurthy for introducing me to the Leslie Orgel quote at the beginning of the introduction. I am thankful to Longfei Wu, Craig Walton, Oliver Shorttle, John Sutherland, and many others connected with the Cambridge Universal Life Initiative for many helpful conversations and new ideas. This work is funded by the Simons Foundation (SCOL Grant \#599634).

\backmatter

\renewcommand\bibname{References}

\end{document}